# The TES-based Cryogenic AntiCoincidence Detector (CryoAC) of ATHENA X-IFU: a large area silicon microcalorimeter for background particles detection

M. D'Andrea [*,1] • C. Macculi[1] • S. Lotti[1] • L. Piro[1] • A. Argan[2] •
G. Minervini[2] • G. Torrioli[3] • F. Chiarello[3] • L. Ferrari Barusso[4] •
E. Celasco[4,5] • G. Gallucci[4,5] • F. Gatti[4,5] • D. Grosso[4,5] •
M. Rigano[4,5] • D. Brienza[6] • E. Cavazzuti[6] • A. Volpe[6]

[*]Corresponding Author, matteo.dandrea@inaf.it
[1]INAF/IAPS, Via del Fosso del Cavaliere 100, 00133 Roma, Italy
[2]INAF Headquarters, Viale del Parco Mellini 84, 00136 Roma, Italy
[3]CNR/IFN Roma, Via del Fosso del Cavaliere 100, 00133 Roma, Italy
[4]University of Genoa, Via Dodecaneso 33, 16146 Genoa, Italy
[5]INFN Genoa, Via Dodecaneso 33, 16146 Genoa, Italy
[6]ASI, Via del Politecnico snc, 00133 Roma, Italy

**Abstract** We are developing the Cryogenic AntiCoincidence detector (CryoAC) of the ATHENA X-IFU spectrometer. It is a TES-based particle detector aimed to reduce the background of the instrument. Here, we present the result obtained with the last CryoAC single-pixel prototype. It is based on a 1 cm$^2$ silicon absorber sensed by a single 2mm x 1mm Ir/Au TES, featuring an on-chip heater for calibration and diagnostic purposes. We have illuminated the sample with $^{55}$Fe (6 keV line) and $^{241}$Am (60 keV line) radioactive sources, thus studying the detector response and the heater calibration accuracy at low energy. Furthermore, we have operated the sample in combination with a past-generation CryoAC prototype. Here, by analyzing the coincident detections between the two detectors, we have been able to characterize the background spectrum of the laboratory environment and disentangle the primary (i.e. cosmic muons) and secondaries (mostly secondary photons and electrons) signatures in the spectral shape.

**1 Introduction**

ATHENA is the next ESA flagship X-ray observatory (0.2 – 12 keV), aimed at studying the "Hot and Energetic Universe" [1]. The mission has recently been redefined, and the launch is planned in late 2030s. The X-IFU is one of the two ATHENA focal plane instruments [2]. It is a cryogenic spectrometer based on a large array of TES microcalorimeters (kilo-pixels order), able to perform spatially-resolved high resolution spectroscopy.

The TES array alone is not able to distinguish between target X-ray photons and background particles depositing energy in the detector band, thus limiting the instrument sensitivity. For this reason, the X-IFU hosts a Cryogenic AntiCoincidence detector (CryoAC [3]) to veto such a background. It is TES-based detector, placed < 1 mm underneath the TES array. While X-ray photons are absorbed in the TES array, background particles deposit energy in both detectors, producing a coincidence signal that allows vetoing these unwanted events. A schematics of the working principle is reported in Fig. 1. Accurate Monte Carlo simulations show that the CryoAC allows to reduce the X-IFU particle background by a factor ~ 50, enabling a significant part of the mission science [4][5].

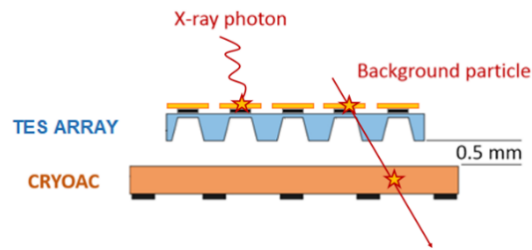

**Fig. 1** Sketch of the CryoAC working principle (detector section view). While X-ray photons are absorbed by TES array pixels, background particles deposit energy on both detectors, producing a coincident signal. (Color figure online)

The CryoAC is a 4-pixels detector, based on large area silicon absorbers sensed by Ir/Au TES. It is required to have wide energy bandwidth, with a low-energy threshold < 20 keV and a saturation energy > 1 MeV. This allows to sense the energy losses of background particles in the absorber, distributed around the most-probable MIP (Minimum Ionizing Particle) energy deposition at ~ 150 keV. Furthermore, it shall have low dead-time (< 1%) and good time-tagging accuracy (10 μs at 1σ), and match strict mechanical, thermal and electromagnetic constraints to ensure the compatibility with the TES array.

## ATHENA X-IFU Cryogenic Anticoincidence Detector (CryoAC)

The last milestone in the CryoAC development has been the production of a single-pixel Demonstration Model (DM) [6]. It has been fully tested in both stand-alone [7] and integrated configuration with TES-array [8], demonstrating full compliance with its requirements. The CryoAC DM design was optimized for athermal phonons collection [9], in order to speed-up the detector at the expense of its spectroscopic response. This has been done by using 96 small TES in-parallel connected, homogeneously covering the absorber surface.

Recently, to test the pure thermal response of the detector without being affected by athermal phonons, we have produced a DM-like prototype featuring a single wide TES. Here below, we report the description of this sample and its testing activity.

## 2 DM127: a CryoAC DM-like thermal prototype

The CryoAC DM-like thermal prototype, namely DM127, is a single-pixel detector based on a 1 cm$^2$ silicon absorber (500 μm thick). As in the CryoAC DM, the absorber is connected to a gold plated-rim via four narrow silicon bridges, achieving a suspended structure with a well-reproducible thermal conductance towards the thermal bath (G ~ 25 nW/K at 100 mK).

The absorber is sensed by a single 2mm x 1mm Ir/Au TES, having normal resistance $R_N$ ~ 200 mΩ and operated at T ~ 100 mK. The TES area is equivalent to the sum of the 96 TES deposited on the DM. For calibration and diagnostic purposes, the sample features also an on-chip heater, constituted by an Au meander deposited on the absorber. A picture of the prototype is shown in Fig. 2.

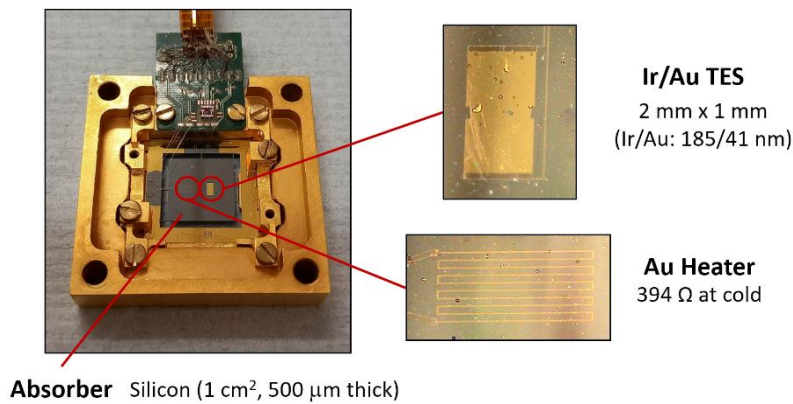

**Fig. 2** CryoAC DM-like thermal prototype, namely DM127 (Color figure online)

The detector assembly includes a Cold Front End Electronics Printed Circuit Board (CFEE PCB) hosting a RuOx thermometer, a shunt resistor for the TES bias ($R_s$ = 8 mΩ) and a single-stage DC-SQUID for its readout. The SQUID is a model M4A produced by VTT [10][11], and it has been operated in the standard Flux Locked Loop (FLL) configuration by a commercial Magnicon XXF-1 electronics[1].

The assembly has been integrated on the mixing chamber of a dilution refrigerator (typical thermal stability ~ 10 $\mu K_{RMS}$ at 50 mK), and magnetically shielded at cold by a Cryophy shield (shielding factor S ~ 500 [12]).

## 3 $^{55}$Fe illumination (6 keV photons)

The detector has been illuminated with a $^{55}$Fe (6 keV photons) radioactive source, at a count rate of ~ 2 cts/s. In order to study the pure thermal detector response, the source has been collimated in a 1 mm diameter spot on the back of the absorber, in a corner distant from the TES.

The sample has been operated at a thermal bath temperature $T_B$ = 50 mK, with a total power dissipation at cold $P_{TOT}$ ~ 10 nW (including TES, shunt resistor and SQUID contributions) and a bias current $I_B$ = 1 mA, in agreement with the CryoAC single-pixel allocations. The working point has been selected to optimize the signal-to-noise ratio of the detected pulses.

Fig. 3 shows one of the acquired pulses and the final energy spectrum obtained by an optimal filtering analysis (on ~ 6000 events). The measured detector low energy threshold is $E_{THR}$ = 1.4 keV (5σ noise level), fully compliant with the CryoAC requirement ($E_{THR}$ < 20 keV, with a < 8 keV goal). The energy spectrum clearly shows both the expected Mn-Kα (5.89 keV) and the Mn-Kβ (6.49 keV) lines, with an energy resolution $\Delta E_{FWHM}$ = (173 ± 1) eV at 6 keV. This is the best energy resolution ever recorded for a CryoAC prototype. Despite there is not a requirement on the CryoAC spectral capabilities (only a goal $\Delta E_{FWHM}$ < 2keV at 6 keV to enable hard-X observations [13]), this is a demonstration that the prototype is well-working from the thermal point of view.

Following this measurement, the sample has been illuminated by collimating the $^{55}$Fe source in different absorber regions, to probe its response from hitting points close and distant from the TES. The results of this activity, aimed to understand the detector athermal/thermal dynamic, is reported in [3].

---

[1] http://www.magnicon.com/squid-electronics/xxf-1

**ATHENA X-IFU Cryogenic Anticoincidence Detector (CryoAC)**

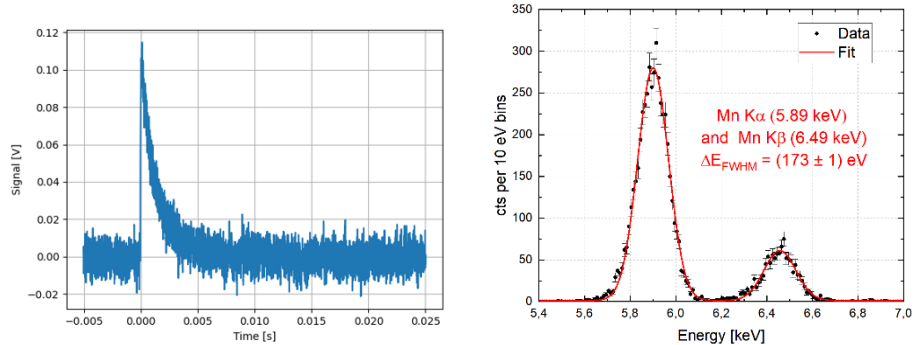

**Fig. 3** Detector response to $^{55}$Fe (6 keV) illumination. *Left* Detected 5.89 keV pulse, raw data. *Right* Acquired energy spectrum obtained by an optimal filtering analysis. Data have been fitted with a double Gaussian distribution. (Color figure online)

## 4 $^{241}$Am illumination (60 keV photons)

The detector has been then illuminated (back side) with a α-shielded $^{241}$Am source. The source has been further shielded with a 1 mm gold-plated copper plate in order to provide only 60 keV photons, at ~ 2 cts/s rate.
The acquired energy spectrum is in good agreement with the expectations (Fig.4 *Left*). The 59.5 keV photopeak is clearly visible, and the detector shows a good spectroscopic response up to that energy, with a measured resolution $\Delta E_{FWHM} = (895 \pm 25)$ eV at 60 keV. At low energy, the spectrum is dominated by the Compton continuum, with the Compton edge at the expected energy for 60 keV photons ($E_{CE, 60\ keV} = 11.3$ keV). Furthermore, it is possible to identify two secondary fluorescences lines, due to the copper (Cu kα at 8.0 keV) and the gold (Au kα at 9.7 keV) generated by the cold stage setup.

To calibrate the detector energy scale on its whole wide bandwidth, the on-chip heater has been used. The sample has been stimulated by injecting fast current pulses (1 μs duration) on the heater, and the detector response pulses have been analyzed by the optimal filtering procedure. This allows to obtain a calibration curve "Optimal filter output" vs "Injected thermal energy" shown in Fig. 4 *Right*. For the heater, energy values have been calculated integrating the heater power over the injected current pulses, without any additional scaling factor. The measurements with $^{55}$Fe and $^{241}$Am (black and red points in Fig. 4 *Right* insets) allowed us to check the calibration accuracy, which resulted to be better than 3% up to 60 keV.

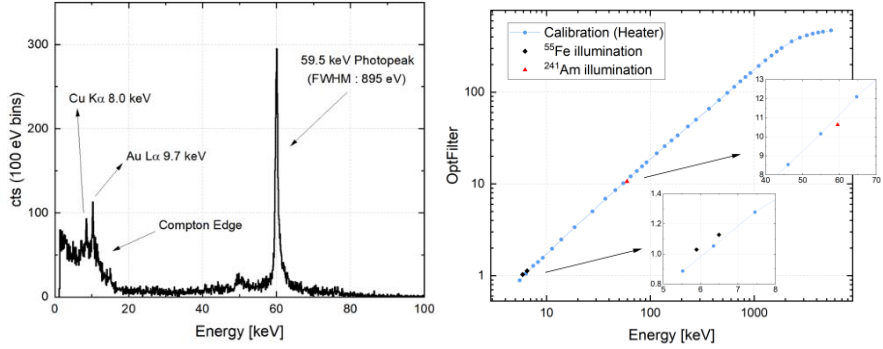

**Fig. 4** *Left* Acquired energy spectrum (obtained by optimal filtering analysis) after the sample illumination by a shielded $^{241}$Am radioactive source (60 keV photons). *Right* Calibration curve of the detector obtained by the on-chip heater (blue points). Reference points obtained with the illumination by $^{55}$Fe (6 keV, black points) and $^{241}$Am (60 keV, red point) are overplotted, and zoomed in the insets. Error bars are inside the point size (Color figure online)

**4 Coincidences measurements**

After the test with radioactive sources, we have performed laboratory background measurements by operating the detector in combination with a past-generation CryoAC prototype (namely AC-S9). This is a DM-like sample that has not been etched to obtain the suspended absorber, thus featuring a large active area (16.6 x 16.6 mm) and a strong connection to the thermal bath (the thermal conductance is defined by gold bonding wires). More detail on this sample can be found in [14].

The simultaneous operation of two detectors allows us to disentangle the background contribution due to soft-X photons (which are mainly absorbed by only one detector) to the one due to cosmic muons (which, depositing energy on both detectors, generate coincident signals). The idea is similar to the CryoAC working principle (Fig. 1).
The measurements have been performed in two different cold-stage configurations, with the detectors in vertical and horizontal alignment with respect to the zenith. This has been done to highlight the anisotropic flux of cosmic muons, which is maximum from the zenith direction [15]. In both configurations the detectors absorbers are 3 mm distant from each other, with empty space in between. The setup picture and schematics are reported in Fig. 5, while some of the detected coincident pulses are shown in Fig. 6.

# ATHENA X-IFU Cryogenic Anticoincidence Detector (CryoAC)

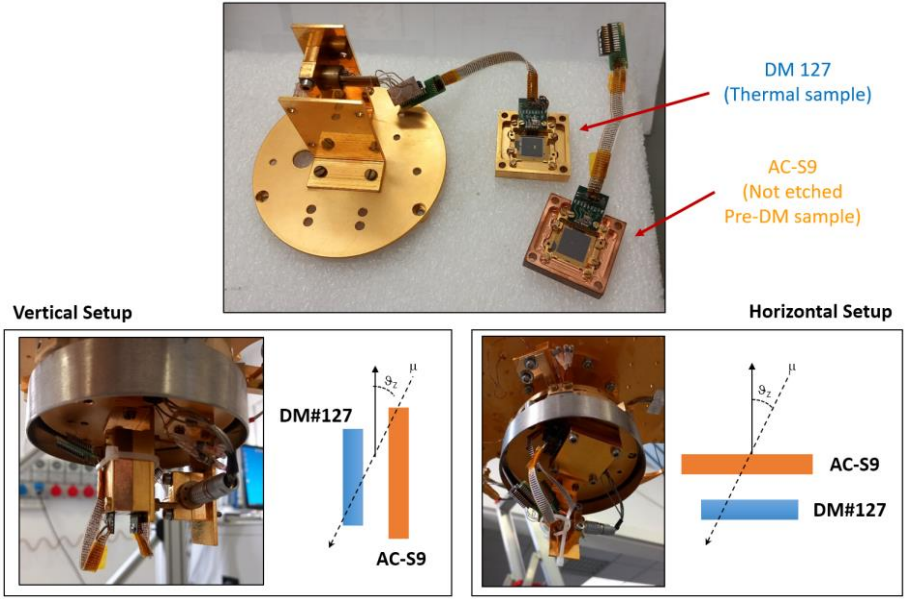

**Fig. 5** *Top* Picture of DM127 and AC-S9 during the integration in the cold stage. *Bottom Left* Picture and schematics of the setup in vertical configuration *Bottom Right* Picture and schematics of the setup in horizontal configuration. (Color figure online)

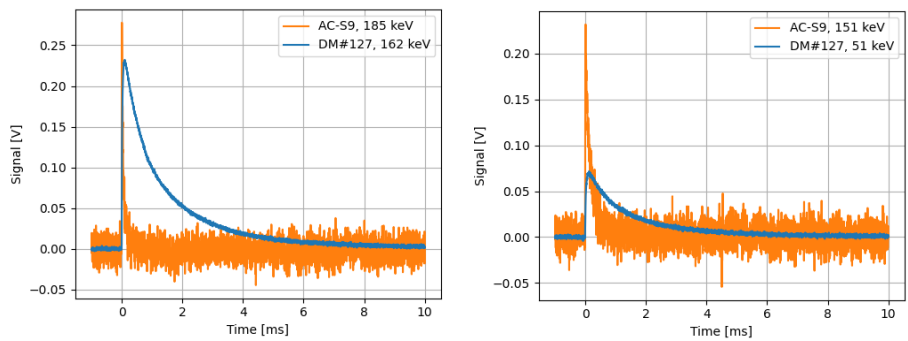

**Fig. 6** Events detected simultaneously (i.e. in coincidence) by DM127 (blue lines) and AC-S9 (orange lines). The AC-S9 pulses are faster due to the high thermal conductance of the not-etched sample. (Color figure online)

The results of the long measurements performed (316 ks observation time in vertical configuration, 530 ks in horizontal configuration) are summarized in Fig. 7, which shows the energy spectra acquired by DM127 in the different setup configurations. In Fig. 7 *Left* the spectra refer to all the detected events, while in Fig. 7 *Right* only to the events that generated also a coincident signal on the second detector (AC-S9).

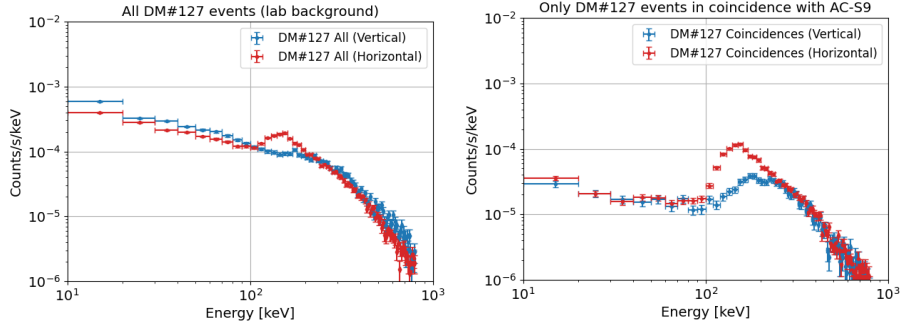

**Fig. 7** Energy spectra of the events detected by DM127 in the different setup configurations (vertical and horizontal). *Left* Energy spectra of all the detected events. *Right* Energy spectra only of the events that generated a coincident signal on the second detector (AC-S9). (Color figure online)

The results are in good agreement with the theoretical expectations. The coincidence measurements (Fig. 7 *Right*) allows indeed to select the background component due to cosmic muons, recognizable by the Landau shape of the spectra. The peak is around 150 keV, as the most probable energy deposition expected for Minimum Ionizing Particles (MIP) in 500 μm of silicon. Furthermore, the higher peak detected in the horizontal configuration is an evidence of the anisotropic flux of the incoming muons. The low energy tail of the spectra is instead probably due to the secondary particles (mainly electrons) generated by the interaction of the cosmic muon with the cryostat and detector surrounding. A more accurate comparison between these results and the expectations is given in another paper of this issue [5], where it is reported the outcome of a Monte Carlo simulation of the experimental setup. Finally, note that the spectra of all the detected events (Fig. 7 *Left*) show an additional component at low energies, characterized by a power-law spectrum and probably due to the X and γ ray background in the laboratory location.

## 5 Conclusions

We have reported the test activity performed on a single-TES CryoAC DM-like sample, namely DM127, produced to work in almost pure thermal regime (without optimization of the athermal phonons collection). The results have been quite encouraging. The detector shows an excellent low energy threshold ($E_{THR}$ = 1.4 keV, requirement at 20 keV) and good spectroscopic capabilities up to 60 keV ($\Delta E_{FWHM}$ = 173 eV at 6 keV and $\Delta E_{FWHM}$ = 895 eV at 60 keV). The on-chip heater allows to perform an energy-scale calibration of the detector on its whole energy bandwidth (up to ~ 1 MeV), with a measured accuracy ~3% at low energy. Finally, working in combination with a past-generation CryoAC prototype, the sample has been able to detect background spectra of the laboratory environment in agreement with the expectations.

Further test activity performed on this sample, mainly related to the investigation of the detector athermal/thermal dynamic and its timing performance, are reported in another paper ([3]). We just highlight that, given the measured value of the thermal time constants and the expected in orbit count-rate, by applying the trigger logic algorithm implemented in the CryoAC Warm Back End Electronics (WBEE) we have verified the detector compliance with both the deadtime and time-tagging accuracy requirements ([3][16]).

All these results could support a change of baseline in the CryoAC configuration, moving from an "athermal" to a "thermal" optimized design. This solution, not requiring a multi-TES network, has the advantage to simplify the samples production, thus providing a more robust baseline. A dedicated trade-off study will be carried out in the coming months, in the context of the ATHENA X-IFU reformulation activity.

**Acknowledgements** This work has been supported by ASI (Italian Space Agency) contract n. 2019-27-HH.0.